\PassOptionsToPackage{table}{xcolor}
\documentclass[sigconf]{acmart}

\usepackage{booktabs} %
\usepackage{array,multirow,graphicx,float}
\usepackage{afterpage}
\usepackage{xcolor}
\usepackage[colorinlistoftodos,prependcaption]{todonotes}
\usepackage[multiple]{footmisc}
\usepackage[capitalize,nameinlink,noabbrev]{cleveref}
\usepackage[utf8]{inputenc}
\let\svtodo\todo\renewcommand\todo[1]{\svtodo[inline]{#1}}

\setcopyright{rightsretained}
\acmDOI{XXX}

\acmISBN{XXX}

\acmConference[XXX]{XXX}{XXX}{XXX}
\acmYear{XXX}
\copyrightyear{2018}

\acmArticle{4}
\acmPrice{XX}

\begin{document}
\title[Embedding Complexity in the Representation]{Embedding Complexity In the Data Representation\\ Instead of In the Model}
\subtitle{A Case Study Using Heterogeneous Medical Data}

\author{Jacek M. Bajor}
\affiliation{%
  \institution{Vanderbilt University Medical Center}
  \city{Nashville}
  \state{Tennessee}
  \country{USA}
}
\email{jacek.m.bajor@vanderbilt.edu}

\author{Diego A. Mesa}
\affiliation{%
  \institution{Vanderbilt University Medical Center}
  \city{Nashville}
  \state{Tennessee}
  \country{USA}
}
\email{diego.mesa@vanderbilt.edu}

\author{Travis J. Osterman}
\affiliation{%
  \institution{Vanderbilt University Medical Center}
  \city{Nashville}
  \state{Tennessee}
  \country{USA}
}
\email{travis.osterman@vanderbilt.edu}

\author{Thomas A. Lasko}
\affiliation{%
  \institution{Vanderbilt University Medical Center}
  \city{Nashville}
  \state{Tennessee}
  \country{USA}
}
\email{tom.lasko@vanderbilt.edu}

\renewcommand{\shortauthors}{J. Bajor et al.}

\begin{abstract}
  Electronic Health Records have become popular sources of data for secondary
  research, but their use is hampered by the amount of effort it takes to
  overcome the sparsity, irregularity, and noise that they contain. Modern
  learning architectures can remove the need for expert-driven feature
  engineering, but not the need for expert-driven preprocessing to abstract
  away the inherent messiness of clinical data. This preprocessing effort is
  often the dominant component of a typical clinical prediction project.

  In this work we propose using semantic embedding methods to directly couple
  the raw, messy clinical data to downstream learning architectures with truly
  minimal preprocessing. We examine this step from the perspective of capturing
  and encoding complex data dependencies in the data representation instead of
  in the model, which has the nice benefit of allowing downstream processing to
  be done with fast, lightweight, and simple models accessible to researchers
  without machine learning expertise. We demonstrate with three typical
  clinical prediction tasks that the highly compressed, embedded data
  representations capture a large amount of useful complexity, although in some
  cases the compression is not completely lossless.
\end{abstract}

\ccsdesc[500]{Applied computing~Health informatics}
\ccsdesc[300]{Computing methodologies~Learning latent representations}

\keywords{Electronic Health Records, Representation Learning, Semantic
  Embedding, Data Representation}

\maketitle

\section{Introduction}
\label{sec:introduction}

An Electronic Health Record (EHR) is a complex collection of heterogeneous data
representing many different types of observations that occur in the course of
medical care. A complete EHR typically contains demographic information,
textual clinical notes, clinical images, medication exposures, laboratory test
results, billing codes, and administrative data such as appointment and
encounter times. Most of these data are sequential and longitudinal, meaning
that observations of a given variable happen repeatedly over the course of a
patient's history, but these observations occur sparsely, irregularly, and
asynchronously with respect to other variables. In addition, the total number
of variables that could be observed at a given time is in the tens of
thousands, without counting the dimensionality of text or images. These
properties present nontrivial challenges to downstream analysis
\cite{Safran2007,Weiskopf2013b}.

Despite these challenges, the secondary use of EHR data for research purposes
has blossomed in the past decade, due to their advantages over randomized
controlled trials or cohort studies \cite{Bowton2014,Wilke2011,Kohane2011}.

Traditionally, the messiness of EHR data was overcome during feature
engineering, in which a domain expert would design features that were not only
informative for the learning task, but that also abstracted away the problems
of raw clinical data. The emergence of deep architectures and other methods
that can learn predictive features directly from data has reduced the need for
much of this engineering, but even these powerful methods do not easily
overcome the messiness of clinical data, which still requires substantial
domain expert knowledge and computational effort to preprocess into a substrate
suitable for learning \cite{Ching2018,Miotto2016,Rajkomar2018}.

Therefore, we would like a way to overcome the need for expensive,
expert-driven preprocessing in the same way that deep architectures overcome
the need for expensive, expert-driven feature engineering. In this work, we
propose the use of full-record semantic embedding methods
\cite{Mikolov2013,LeMikolov2014} to directly couple messy clinical data to
downstream learning architectures. In addition to removing the need for heavy
preprocessing, the embeddings provide the nice benefit that many of the complex
data relationships that would have been encoded in the model are instead
encoded in the data representation. This allows them to be used with simple
linear models that are not only computationally cheaper, but also more
accessible to researchers without machine learning expertise.

\subsection{Previous Work}
\label{subsec:previous_work}

Recently, much work has been done on developing compact and functional
representations of medical records, including the use of deep learning over EHR
data \cite{shickel2017deep}.

A notable attempt is the Deep Patient framework of Miotto and colleagues
\cite{Miotto2016}, which uses vector representations of patient records
generated by stacked denoising autoencoders. The approach started with
bag-of-words counts of clinical codes and concepts extracted from textual
notes, and then used Latent Dirichlet Allocation to compress the large
resulting vectors into a more manageable representation for the autoencoders.

Beaulieu-Jones and colleagues \cite{BeaulieuJones2016} also used denoising
autoencoders to develop their patient representation from various binary
clinical descriptors, although their assessment was only on synthetic data.

Other work has used semantic embedding at less than full-record scope to build
aspects of patient representations. Choi and colleagues \cite{ChoiSSS16} summed
word-level skip-gram embedded vectors of clinical codes to create a full-record
representation \cite{Mikolov2013}.  Choi and colleagues \cite{ChoiBSCS16} used
a multi-level embedding model that represents a single patient visit as a
skip-gram-type embedding of precomputed word-level code embeddings similar to
skip-gram vectors but constrained to have nonnegative values for
interpretability.  Pham and colleagues \cite{pham2016deepcare} generated
word-level semantic embeddings for diagnosis and intervention codes, using
pooling and concatenation to aggregate them into a vector representing a single
admission. Nguyen and colleagues \cite{nguyen2017mathtt} used word-level
embedding as preprocessing for a Convolutional Neural Network architecture.

\begin{figure}[t]
  \includegraphics[width=3.33in]{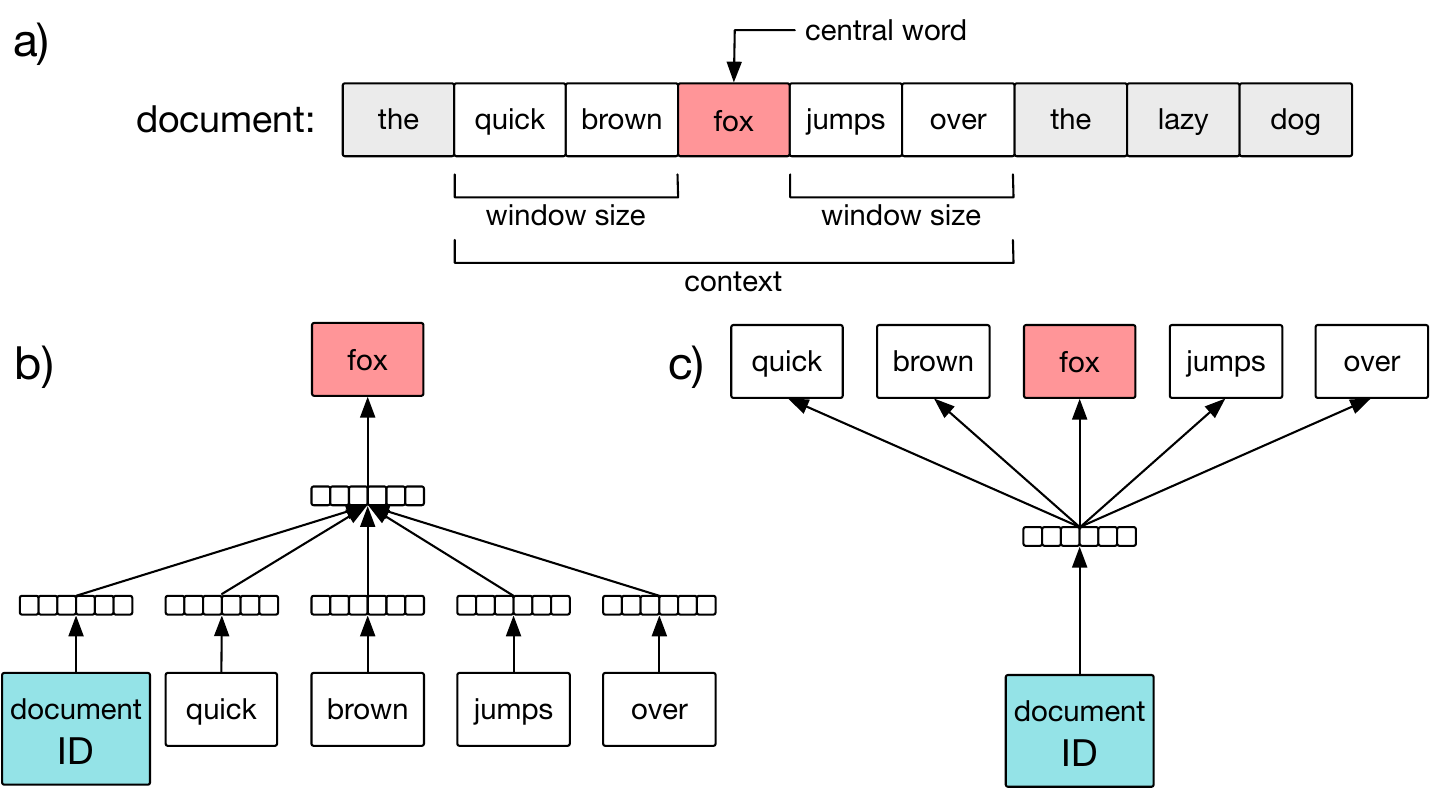}
  \caption{Architecture of the semantic embedding learning tasks, explained
    using a text document metaphor. a) A sample document, indicating the
    central word in red and the context words in white, which are defined
    by the window size. b) The Distributed Memory Model, in which the target
    word is predicted given the vectors of the document and context words. The
    lines of small squares represents the learned semantic vectors. c) The
    Distributed Bag of Words Model, in which the target and context words are
    predicted given the document's semantic vector. Figure adapted from Le et
    al. \cite{LeMikolov2014}.}
  \label{doc2vec-figure}
\end{figure}

Other efforts have aimed at encoding temporal aspects of EHR data in the
predictive model.  Choi and colleagues \cite{choi2016doctor} used time-stamped
events as inputs to a particular type of Recurrent Neural Network (RNN) to
predict future disease diagnosis.  Mehrabi and colleagues
\cite{mehrabi2015temporal} constructed straightforward temporal matrix
representations using codes in rows and years in columns as inputs to a deep
Boltzmann machine.

Another interesting recent effort is Rajkomar and colleagues' \cite{Rajkomar2018}
mapping of raw EHR data to the Fast Healthcare Interoperability Resources
(FHIR) format \footnote{http://hl7.org/fhir} to encode EHR information for several different
sequence-oriented models.

All of these approaches rely on preprocessing schemes to prepare the data for
use in a model --- schemes that are in some cases quite elaborate, require
expert tuning or are unique to
a specific EHR structure. They can be difficult to compute end-to-end and
expensive to train, requiring significant amounts of time and computational
resources. In contrast, our approach requires truly minimal preprocessing or
tuning, which makes it easily generalizable between institutions.

\subsection{Main Contribution}
\label{subsec:main_contribution}

In this paper we propose using a full-record semantic embedding to represent a
patient's entire medical history in a compact but expressive form. The method
uses an established embedding algorithm that is relatively easy to implement.
It does not rely on extensive data engineering or preprocessing, but takes as
input commonly used time-stamped data, making it useful for a wide range
of data sources.  We show that this representation can be used in several
typical medical prediction problems under simple linear models, saving both
time and computational resources. This representation is particularly powerful
for rapidly testing secondary-use research ideas with EHR data, because it can
be precomputed for all patients and then used in downstream modeling by
non-experts.

Full source code for the project, including all steps from data
extraction through model training, evaluation, and figure generation is
publicly available from our Github
repository\footnote{https://github.com/ComputationalMedicineLab/patient2vec}.
The datasets themselves cannot be publicly released due to the sensitive nature
of medical data.

\section{Background}
\label{sec:background}

\subsection{Medical Taxonomies}
\label{subsec:medical_concept_classifications}

\begin{figure*}[t]
  \includegraphics[width=\textwidth]{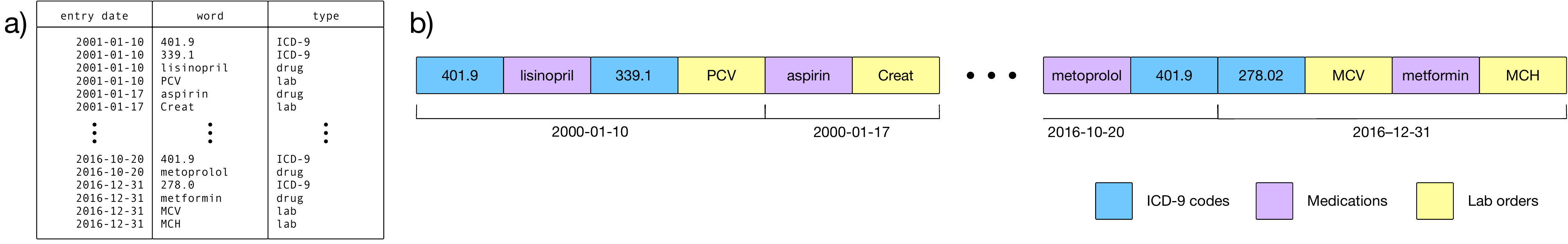}
  \caption{Example input data for one patient record. a) Raw data as obtained
    from the EHR. b) Data after conversion to a chronological sequence for
    model training, where events within the same day are ordered randomly.}
  \label{input-data-figure}
\end{figure*}

There are several taxonomies that encode variables of interest in an EHR,
including diagnosis and procedure codes, medications names, and laboratory test
names, although many of them are not universally adopted. In this section, we
briefly describe the taxonomies used in our project.

\paragraph{ICD-9 Codes} Whenever a patient has billable contact with the
healthcare system, date-stamped diagnosis codes are attached to the record,
indicating the medical conditions that were relevant to the encounter. These
codes are notoriously unreliable because, among other things, they are often
assigned as a guess before the final diagnosis is known, and they are not
revised later \cite{OMalley2005}. This works fine for billing purposes, but
causes obvious trouble if we don't allow for a high level of noise in these
codes. However,  when used in aggregate \cite{Denny2010}, and especially if used
probabilistically \cite{Lasko2014}, they can be a valuable source of disease
signals.

In our institution, codes from the International Classification of Diseases,
Ninth Revision (ICD-9)\footnote{https://www.cdc.gov/nchs/icd/} have
historically been used for diagnostic codes, although the Tenth Revision
(ICD-10) has recently been adopted. We used only the ICD-9 version in this
project.

The ICD-9 hierarchy consists of 21 chapters, each roughly corresponding to a
single organ system or pathologic class. Within a chapter, three-digit parent
codes indicate a general disease area (such as female breast cancer of any
type), and leaf-level codes of up to five digits indicate specialized
distinctions within that area. A small number of ICD-9 codes represent medical
procedures. The full classification contains over 18,000 unique codes. We used
leaf-level codes as part of the input to our embedding models, and parent codes
in evaluation cohort definitions.

\paragraph{Phecodes} Because the vocabulary of ICD-9 or other taxonomies is so
large, and the distinctions they encode are not always important for research
purposes, equivalence classes have been defined to group them into a smaller
number of codes. The phecode taxonomy is one such grouper that maps ICD-9 codes
down to 1,866 phecodes at leaf level, with each phecode representing a common
medical condition \cite{Denny2010,Wei2017a}. We did not use this grouper when
training our embedded representation, but we did use it to reduce the
dimensionality of the comparison representation for computational tractability
(\cref{paragraph:traditional_representation}).

\paragraph{ATC codes} The Anatomical Chemical Classification
System\footnote{http://www.whocc.no/atc/structure\_and\_principles/} (ATC) is a
multi-level grouper for medications, organized by both anatomic and therapeutic
class.  As with the phecode grouper, we did not use the ATC grouper to train
our embedded representation, but we did use it to reduce the dimensionality of
our comparison representation for computational tractability
(\cref{paragraph:traditional_representation}).

\subsection{Semantic embeddings}
\label{subsec:semantic_embeddings}

A semantic embedding is a vector representation of a set of variables that
attempts to encode the semantic meaning of each variable in a way that is
accessible to downstream computing. Its first demonstration was in learning
semantic vectors of words in a document such that words of similar meaning were
located near each other in the embedded vector space, and that the relative
location of two words in the space could encode a meaningful relationship, such
as the relationship of a capital city to a country \cite{Mikolov2013}.

Two algorithms that were initially proposed were the Continuous Skip-gram
Model, in which the learning problem was to predict nearby words (called
\emph{context words}) using the central word's semantic vector, and the
Continuous Bag of Words Model, in which the problem was to predict the central
word given the semantic vectors of the context words \cite{Mikolov2013}.

\begin{table*}[t]
  \resizebox{\textwidth}{!}{
    \begin{tabular}{rccccccc}
      \hline
      & \multicolumn{1}{c}{\textbf{Embedding}} & \multicolumn{2}{c}{\textbf{Breast cancer}} & \multicolumn{2}{c}{\textbf{Diabetes treatment}} & \multicolumn{2}{c}{\textbf{Lung cancer}} \\
      & & \multicolumn{1}{c}{\textbf{Positive}} & \multicolumn{1}{c}{\textbf{Negative}} & \multicolumn{1}{c}{\textbf{Positive}} & \multicolumn{1}{c}{\textbf{Negative}} & \multicolumn{1}{c}{\textbf{Positive}} & \multicolumn{1}{c}{\textbf{Negative}} \\
      \hline
      \# of ICD-9 events & 79,866,333 & 300,226 & 300,248 & 683,538 & 683,542 & 152,331 & 1,059,198 \\
      \# of lab events & 216,392,248 & 710,161 & 741,293 & 1,476,329 & 1,554,251 & 445,196 & 3,626,106 \\
      \# of medication events & 66,269,824 & 266,289 & 271,989 & 467,408 & 461,650 & 147,244 & 1,118,321 \\
      \# of total events & 362,528,405 & 1,276,676 & 1,313,530 & 2,627,275 & 2,699,443 & 744,771 & 5,803,625 \\
      \hline
      \# of unique ICD-9 & 19,994 & 7,094 & 8,261 & 9,459 & 10,933 & 5,600 & 10,720 \\
      \# of unique labs & 5,509 & 1,240 & 1,510 & 1,643 & 1,949 & 1,038 & 2,117 \\
      \# of unique codes & 31,589 & 9,671 & 11,348 & 12,625 & 14,756 & 7,694 & 14,829 \\
      \hline
      \# of patients & 2,309,712 & 2,901 & 2,901 & 10,477 & 10,477 & 1,104 & 5,631 \\
      months of history & 11.8 [0, 67.3] & 90.1 [55.6, 132.5] & 143.2 [100.9, 186.7] & 70 [42.7, 110] & 129.1 [87.5, 173.6] & 80.1 [49.9, 124.9] & 71.6 [43.3, 114.2] \\
      ICD-9 events & 8 [3, 27] & 51 [25, 122] & 51 [25, 122] & 39 [21, 76] & 39 [21, 76] & 79.5 [33, 174] & 92 [35, 231] \\
      Lab events & 1 [0, 49] & 85 [18, 247] & 78 [13, 261] & 60 [8, 155] & 48 [2, 143] & 183.5 [48.8, 466.2] & 208 [42, 674] \\
      Medication events & 2 [0, 14] & 23 [4, 87] & 26 [7, 88] & 11 [2, 43] & 16 [4, 46] & 49 [10, 157.5] & 54 [9, 201.5] \\
      Total events & 21 [4, 93] & 168 [59, 461] & 163 [64, 475] & 118 [50, 275] & 114 [49, 265] & 329.5 [104.8, 792.8] & 369 [98, 1130.5] \\
      \hline
    \end{tabular}
  }
  \caption{Composition of the data used for embedding and the three evaluation
    problems. Cells contain either total counts or Median [IQR].}
  \label{data-description-table}
\end{table*}

These algorithms were extended to learn a single semantic vector representing
an entire document \cite{LeMikolov2014}.  One extension, the Distributed Memory
Model (\cref{doc2vec-figure}b), is analogous to the Continuous Bag of Words
Model, in which the task is to predict a target word given the vectors of
nearby words and the vector of the document as a whole. The second extension is
the Distributed Bag of Words Model (\cref{doc2vec-figure}c), analogous to the
Skip-gram Model, in which the task is to predict individual document words
given the document vector.

All of these models use a simple neural network architecture in which the
dimension of the input and output layers is the size of the vocabulary, and
once trained, the hidden layer contains the semantic vectors of interest.

Although these architectures are simple, the number of nodes in them can be
very large, which increases training time. To reduce this time, two alternative
training methods are commonly used: hierarchical softmax
\cite{Morin2005HierarchicalPN} and negative sampling \cite{Mnih08ascalable}.
Both increase speed by updating only a fraction of all weights per iteration.
Hierarchical softmax is a computationally efficient approximation of the
softmax function which uses a binary tree representation of all words in the
vocabulary.  The words themselves are leaves in the tree. For each leaf, there
exists a unique path from the root to the leaf, and this path is used to
estimate the probability of the word represented by the leaf
\cite{Colyer}. Negative sampling is a simplified variant of Noise Contrastive
Estimation (NCE) \cite{GutmannHyvarinen2012,MikolovSCCD13}, where only a
sample of output words are updated per iteration. The target output word
is kept in the sample and gets updated, but a number of non-targets are added
as negative samples \cite{Colyer}.

In this project, we used the document-level embedding approach, treating an
entire patient record (which could cover more than 20 years of history) as a
document, and the data elements of ICD-9 codes, lab tests, and medications as
its words.

\section{Patient-Level Embedding Models}
\label{sec:patient_level_embedding_models}

We tested both the Distributed Memory Model and the Distributed Bag of Words
Model to create record-level embeddings, using both hierarchical softmax
and negative sampling approximation methods.

\subsection{Data}
\label{subsec:ple-data}

All data for this project was extracted from the de-identified mirror of
Vanderbilt's Electronic Health Record, which contains administrative data,
billing codes, medication exposures, laboratory test results, and narrative text
for over 2 million patients, reaching back nearly 30 years \cite{Roden2008}. We
obtained IRB approval to use this data in this research.

To train the embedding model, we extracted all ICD-9 billing codes, medication
exposures, and laboratory test results from each patient record
(\cref{data-description-table}). Data preprocessing was deliberately
minimal. ICD-9 code events and generic medication names were used as-is. No attempts at
synonym detection, grouping, or typographical error correction were
made. Laboratory test results were represented only as measurement events,
labeled by test name and time, ignoring the numeric result.  For historical
reasons, many nearly identical laboratory tests were represented by distinct
identifiers, and we did not attempt to group them. The only transformation step
was to remove vocabulary elements appearing less than 250 times in the dataset
(out of 363 million total events), resulting in a final vocabulary size of
31,589.

For each record, these elements were ordered by the sequence of their
appearance (\cref{input-data-figure}). Order within the same day was randomized
because some events included only date information.

A total of 2,309,712 patient records and 362,528,405 events were used to
train the model.

\subsection{Model Training}
\label{subsec:model_training}

We computed an embedded representation for each patient record using both the
Distributed Bag of Words Model and Distributed Memory Model architectures. For
each architecture, all combinations of embedding dimension (10, 50, 100, 300,
500, 1000), sliding window sizes (5, 10, 20, 30, 50), and softmax approximation
methods (hierarchical softmax, negative sampling) were trained and evaluated.

Embedding models were trained for 20 iterations, with convergence usually after
5 - 10 iterations. After initial evaluation (\cref{models-table}), the top 15
performing models were trained for an additional 60 iterations.

All models were generated using Gensim \cite{Rehurek2010}, an open-source
Python library for statistical semantic analysis and natural language
processing.

\section{Evaluation}
\label{sec:model_evaluation}

\begin{table*}[t]
  \resizebox{\textwidth}{!}{
    \begin{tabular}{|c|c|cccccccccccccccccccc|}
    \hline
    \multicolumn{2}{|c}{algorithm} & \multicolumn{10}{|c}{Distributed Memory version of
    Paragraph Vector} & \multicolumn{10}{|c|}{Distributed Bag of Words version of Paragraph Vector} \\
    \hline
    \multicolumn{2}{|c}{softmax} & \multicolumn{5}{|c}{Hierarchical Softmax} & \multicolumn{5}{|c}{Negative Sampling} & \multicolumn{5}{|c}{Hierarchical Softmax} & \multicolumn{5}{|c|}{Negative Sampling}\\
    \hline
    \multicolumn{2}{|c}{window size} & \multicolumn{1}{|c}{5} & \multicolumn{1}{|c}{10} & \multicolumn{1}{|c}{20} & \multicolumn{1}{|c}{30} & \multicolumn{1}{|c}{50} &
      \multicolumn{1}{|c}{5} & \multicolumn{1}{|c}{10} & \multicolumn{1}{|c}{20} & \multicolumn{1}{|c}{30} & \multicolumn{1}{|c}{50} &
      \multicolumn{1}{|c}{5} & \multicolumn{1}{|c}{10} & \multicolumn{1}{|c}{20} & \multicolumn{1}{|c}{30} & \multicolumn{1}{|c}{50} &
      \multicolumn{1}{|c}{5} & \multicolumn{1}{|c}{10} & \multicolumn{1}{|c}{20} & \multicolumn{1}{|c}{30} & \multicolumn{1}{|c|}{50}\\
    \hline
    \parbox[t]{2mm}{\multirow{6}{*}{\rotatebox[origin=c]{90}{embedding size}}}
    & 10   & \cellcolor[HTML]{228d8d} \textcolor{white}{0.71} & \cellcolor[HTML]{26828e} \textcolor{white}{0.70} & \cellcolor[HTML]{433e85} \textcolor{white}{0.64} & \cellcolor[HTML]{440256} \textcolor{white}{0.60} & \cellcolor[HTML]{471063} \textcolor{white}{0.61} & \cellcolor[HTML]{1f968b} \textcolor{white}{0.72} & \cellcolor[HTML]{31678e} \textcolor{white}{0.68} & \cellcolor[HTML]{3a538b} \textcolor{white}{0.66} & \cellcolor[HTML]{472d7b} \textcolor{white}{0.63} & \cellcolor[HTML]{440154} \textcolor{white}{0.60} & \cellcolor[HTML]{77d153} 0.79 & \cellcolor[HTML]{6ccd5a} 0.78 & \cellcolor[HTML]{77d153} 0.79 & \cellcolor[HTML]{77d153} 0.79 & \cellcolor[HTML]{70cf57} 0.78 & \cellcolor[HTML]{89d548} 0.79 & \cellcolor[HTML]{65cb5e} 0.78 & \cellcolor[HTML]{a8db34} 0.81 & \cellcolor[HTML]{84d44b} 0.79 & \cellcolor[HTML]{bade28} 0.81\\
    & 50   & \cellcolor[HTML]{75d054} 0.79 & \cellcolor[HTML]{32b67a} \textcolor{white}{0.75} & \cellcolor[HTML]{228c8d} \textcolor{white}{0.71} & \cellcolor[HTML]{277e8e} \textcolor{white}{0.70} & \cellcolor[HTML]{3a538b} \textcolor{white}{0.66} & \cellcolor[HTML]{27ad81} \textcolor{white}{0.75} & \cellcolor[HTML]{1f9e89} \textcolor{white}{0.73} & \cellcolor[HTML]{2b748e} \textcolor{white}{0.69} & \cellcolor[HTML]{39568c} \textcolor{white}{0.66} & \cellcolor[HTML]{365d8d} \textcolor{white}{0.67} & \cellcolor[HTML]{f6e620} 0.83 & \cellcolor[HTML]{d5e21a} 0.82 & \cellcolor[HTML]{c0df25} 0.81 & \cellcolor[HTML]{e5e419} 0.83 & \cellcolor[HTML]{f6e620} 0.83 & \cellcolor[HTML]{a5db36} 0.80 & \cellcolor[HTML]{addc30} 0.81 & \cellcolor[HTML]{c2df23} 0.81 & \cellcolor[HTML]{bade28} 0.81 & \cellcolor[HTML]{b0dd2f} 0.81 \\
    & 100  & \cellcolor[HTML]{90d743} 0.80 & \cellcolor[HTML]{56c667} \textcolor{white}{0.77} & \cellcolor[HTML]{1e9d89} \textcolor{white}{0.73} & \cellcolor[HTML]{20928c} \textcolor{white}{0.72} & \cellcolor[HTML]{32658e} \textcolor{white}{0.67} & \cellcolor[HTML]{58c765} \textcolor{white}{0.77} & \cellcolor[HTML]{22a884} \textcolor{white}{0.74} & \cellcolor[HTML]{25838e} \textcolor{white}{0.70} & \cellcolor[HTML]{238a8d} \textcolor{white}{0.71} & \cellcolor[HTML]{375a8c} \textcolor{white}{0.66} & \cellcolor[HTML]{eae51a} 0.83 & \cellcolor[HTML]{dae319} 0.82 & \cellcolor[HTML]{dfe318} 0.82 & \cellcolor[HTML]{f8e621} 0.83 & \cellcolor[HTML]{eae51a} 0.83 & \cellcolor[HTML]{dae319} 0.82 & \cellcolor[HTML]{e2e418} 0.83 & \cellcolor[HTML]{81d34d} 0.79 & \cellcolor[HTML]{dae319} 0.82 & \cellcolor[HTML]{c0df25} 0.81 \\
    & 300  & \cellcolor[HTML]{8ed645} 0.80 & \cellcolor[HTML]{4cc26c} \textcolor{white}{0.77} & \cellcolor[HTML]{1fa187} \textcolor{white}{0.73} & \cellcolor[HTML]{1f968b} \textcolor{white}{0.72} & \cellcolor[HTML]{238a8d} \textcolor{white}{0.71} & \cellcolor[HTML]{4cc26c} \textcolor{white}{0.77} & \cellcolor[HTML]{1f958b} \textcolor{white}{0.72} & \cellcolor[HTML]{20928c} \textcolor{white}{0.72} & \cellcolor[HTML]{2a778e} \textcolor{white}{0.69} & \cellcolor[HTML]{2c738e} \textcolor{white}{0.69} & \cellcolor[HTML]{d8e219} 0.82 & \cellcolor[HTML]{fde725} 0.84 & \cellcolor[HTML]{fde725} 0.84 & \cellcolor[HTML]{dfe318} 0.82 & \cellcolor[HTML]{dfe318} 0.82 & \cellcolor[HTML]{c5e021} 0.82 & \cellcolor[HTML]{d0e11c} 0.82 & \cellcolor[HTML]{d2e21b} 0.82 & \cellcolor[HTML]{d0e11c} 0.82 & \cellcolor[HTML]{b5de2b} 0.81 \\
    & 500  & \cellcolor[HTML]{8bd646} 0.79 & \cellcolor[HTML]{6ece58} 0.78 & \cellcolor[HTML]{1fa088} \textcolor{white}{0.73} & \cellcolor[HTML]{1f988b} \textcolor{white}{0.72} & \cellcolor[HTML]{26828e} \textcolor{white}{0.70} & \cellcolor[HTML]{60ca60} 0.78 & \cellcolor[HTML]{26ad81} \textcolor{white}{0.74} & \cellcolor[HTML]{1e9d89} \textcolor{white}{0.73} & \cellcolor[HTML]{24868e} \textcolor{white}{0.71} & \cellcolor[HTML]{287d8e} \textcolor{white}{0.70} & \cellcolor[HTML]{cae11f} 0.82 & \cellcolor[HTML]{bade28} 0.81 & \cellcolor[HTML]{d8e219} 0.82 & \cellcolor[HTML]{b5de2b} 0.81 & \cellcolor[HTML]{cde11d} 0.82 & \cellcolor[HTML]{93d741} 0.80 & \cellcolor[HTML]{cde11d} 0.82 & \cellcolor[HTML]{9bd93c} 0.80 & \cellcolor[HTML]{95d840} 0.80 & \cellcolor[HTML]{89d548} 0.79 \\
    & 1000 & \cellcolor[HTML]{7fd34e} 0.79 & \cellcolor[HTML]{46c06f} \textcolor{white}{0.77} & \cellcolor[HTML]{1f9f88} \textcolor{white}{0.73} & \cellcolor[HTML]{20a386} \textcolor{white}{0.74} & \cellcolor[HTML]{228b8d} \textcolor{white}{0.71} & \cellcolor[HTML]{40bd72} \textcolor{white}{0.76} & \cellcolor[HTML]{2db27d} \textcolor{white}{0.75} & \cellcolor[HTML]{2a788e} \textcolor{white}{0.69} & \cellcolor[HTML]{1f988b} \textcolor{white}{0.72} & \cellcolor[HTML]{1f948c} \textcolor{white}{0.72} & \cellcolor[HTML]{bddf26} 0.81 & \cellcolor[HTML]{e5e419} 0.83 & \cellcolor[HTML]{d8e219} 0.82 & \cellcolor[HTML]{cde11d} 0.82 & \cellcolor[HTML]{b8de29} 0.81 & \cellcolor[HTML]{86d549} 0.79 & \cellcolor[HTML]{81d34d} 0.79 & \cellcolor[HTML]{8ed645} 0.80 & \cellcolor[HTML]{9bd93c} 0.80 & \cellcolor[HTML]{b2dd2d} 0.81 \\
    \hline
    \end{tabular}
  }
  \caption{Effects of varying the embedding parameters. The Distributed Bag of
    Words approach is a clear winner in our problems, with performance being
    fairly invariant to other parameters away from the extremes. The number in
    each cell is the AUC of the gradient tree boosting model with default
    parameters predicting breast cancer diagnosis with a 12-month prediction
    horizon (boosting model parameters were optimized subsequent to this
    step). Dark purple color represents the worst performance, bright yellow
    the best.}
  \label{models-table}
\end{table*}

To evaluate the extent to which the embedded representations preserve,
decrease, or augment the predictive value of the original data, we compared the
performance of those representations to that of a simple but common
representation of the same information on three different clinical tasks, each
task being representative of a typical and meaningful predictive problem in the
clinical domain.  For each problem we trained a simple linear model and a
complex nonlinear model, comparing the performance of our embedded
representation with that of a simple but traditional summary
representation. The goal of the evaluation was not to achieve state-of-the-art
performance on each problem, but to gain insight into the tradeoffs of the
embedded representation vs. the simple one.  There are certainly more complex
models (such as sequence-based models
\cite{Bajor2017,Luo2017,Pham2017,Choi2017} or deep feed-forward networks
\cite{Ching2018,Miotto2017,Miotto2016b,Lasko2013}) that could probably squeeze
a little more performance out of the data for each problem, but they would
require much greater computational effort, and using them here would not
appreciably alter our conclusions.

We used the linear model to understand the degree to which the semantic
embedding captures complex, nonlinear dependencies in the data, and the
nonlinear model to understand the degree to which the embedding loses
predictive information. We expect the embedded representation to outperform the
simple representation on the linear model if it managed to capture meaningful
dependencies between elements. We expect it to under-perform the simple
representation on the nonlinear model if the embedding irretrievably loses
predictive information. Finally, we can compare the ability of the embedding to
capture complex dependencies to the ability of the model to capture those
dependencies by comparing the performance of the embedded representation under
the simple model to the simple representation under the complex model.

In addition to this objective evaluation, we also subjectively investigated the
extent to which the embedding captures information about patient state and
disease trajectory by generating a visualization of patients in the embedded
space and the changes that occur as a disease process progresses.

\subsection{Data}
\label{subsec:eval-data}

Data for the evaluations was drawn from the same source as the embedding
models, although it required some additional preprocessing to identify the
proper cohorts and assign labels. For each test problem, input data was
collected from the repository for each patient record up until a given
problem-specific cutoff date (see below). The embedded representation and
simple representation were then computed from that time-truncated data.

\paragraph{Simple Representation}
\label{paragraph:traditional_representation}

The simple representation we used was a counted bag-of-words vector, where a
word in this case is the appearance of a lab test, medication exposure, or
ICD-9 code in the record. This is a very common representation for medical
prediction projects \cite{Liao2010,Carroll2012,Sun2014,Pivovarov2015}. However,
the large vocabulary this produced (31,520 elements) was computationally
intractable for our models, so we reduced it by grouping ICD-9 codes into
phecodes, and by grouping medications to their ATC therapeutic class
equivalence (\cref{subsec:medical_concept_classifications}). The final number
of features after grouping was between 3639 and 4535, after removing features
with 0 total counts in each problem. The sparsity of this representation was
very high, even after grouping.

\paragraph{Embedded Representation}
\label{paragraph:embedded_representation}

Once the semantic embedding models were trained, we projected the
time-truncated test data into the embedded space in the usual way, which is to
run the training algorithm one more iteration with the single addition of the
new instance.

\subsection{Objective Evaluation}
\label{subsec:objective_evaluation}

We objectively evaluated the embedding models by assessing how well they
perform as input representations for three clinical prediction problems,
comparing them to the performance of the same information in the simple
representation.

Each prediction problem was expanded into five tasks with increasingly
difficult \emph{prediction horizons}, or the time between the input data cutoff
date and the event to be predicted. We used prediction horizons of 1 day and 1,
3, 6, and 12 months. For each task we evaluated discrimination using the area
under the Receiver Operating Characteristic Curve (AUC), and calibration using
observed to expected probabilities in bins over the range of prediction.

The linear model was a simple elastic net (Scikit-learn implementation)
\cite{scikit-learn}, with parameters optimized by cross-validation.

The nonlinear model was the XGBoost implementation of gradient tree boosting,
also known as a gradient boosting machine (GBM) \cite{ChenG16}. This model is
extremely effective at extracting meaningful dependencies in the input data, and
often achieves state-of-the art results. In 2015, 17 out of 29 winning solutions
in Kaggle competitions\footnote{https://www.kaggle.com/} used XGBoost
\cite{ChenG16}. Its drawback is that training and parameter optimization
requires considerable computational time.

In our experiments, a grid search over embedding parameters using default
XGBoost parameters was the first pass at optimization (\cref{models-table}),
and then the top 15 embedding parameter settings were chosen and XGBoost
parameters optimized with random search for each of those embeddings.

\begin{figure*}[t]
  \includegraphics[width=7in]{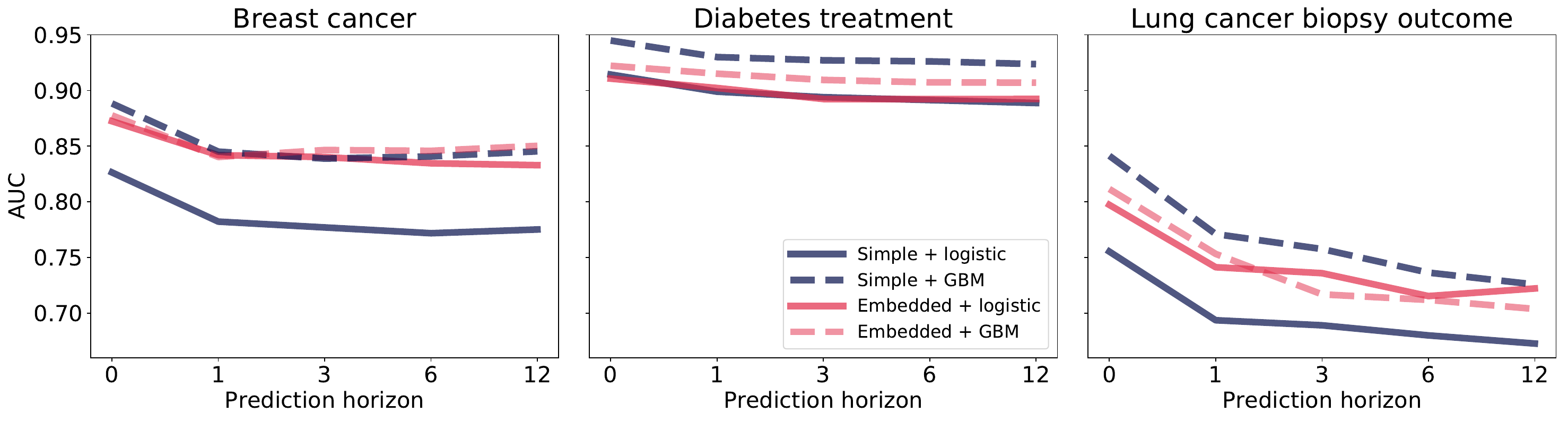}
  \caption{Capture and loss of complex data dependencies. Where the red solid
    line is above the blue solid line, it indicates the capture of complex
    dependencies by the embedded representation. Where the blue dashed line is
    above the red dashed line, it indicates loss of complex information by the
    embedded representation. Where the blue dashed line is above the red solid
    line, it indicates information captured by the complex model that was
    missed by the embedded representation.}
  \label{auc-calibration-figure}
\end{figure*}

\paragraph{Female Breast Cancer Prediction Problem}
\label{paragraph:breast_cancer_prediction}

The first evaluation problem was to predict whether a female patient would
develop breast cancer at the prediction horizon. Female subjects with at least
10 recorded ICD-9 codes of any type and at least 24 months of data before the
cutoff date were considered for the dataset. Records with at least 3 codes with
an ICD-9 174-parent (\emph{Malignant neoplasm of female breast}) were labeled
positive, and their cutoff date was the first of those 174-parent
events. Records with no 174-parent codes at all were labeled negative. Their
cutoff date was set arbitrarily to the day when raw data was pulled from the
database for the experiment.

The raw number of negative instances was much higher than number of positive
instances, but negative instances were then selected to match positive instances
as closely as possible on the number of total ICD-9 codes and the time length of
the record (\cref{data-description-table}). After matching, we had the same
number of negative and positive instances. A stratified split was then performed
to divide the dataset into training (75\%), test (20\%) and validation (5\%)
sets.

\paragraph{Diabetes Treatment Prediction Problem}
\label{paragraph:diabetes_treatment_prediction}

The second evaluation problem was to predict whether a given patient would
begin treatment for type 2 diabetes at the prediction horizon. We used the
start of medical treatment rather than the date of diagnosis because it is a
cleaner event that is easier to identify from the data in the record.  For type
2 diabetes, medical treatment generally begins with an oral glucose-lowering
drug, and we used the start of such a drug to define the prediction target.

All records selected for this task had at least 10 ICD-9 codes of any type and
at least 24 months of data before the cutoff date. Records with at least 10
mentions of a medication in the ATC category A10B: \emph{Blood glucose lowering
drugs, excluding insulins}, and which did not have a prior record of taking
insulin, were labeled as positive instances. The cutoff date was set as the
first mention of such a drug. Records with no mention of any drug in the broader
ATC category A10: \emph{Drugs used in diabetes} were labeled as negative
instances, with a cutoff date arbitrarily set as the day when raw data was
pulled for the experiment.

As above, negative instances were matched to positive instances on the total
number of ICD-9 codes and the time length of the medical record
(\cref{data-description-table}).  A stratified split was performed to
divide the dataset into training (75\%), test (20\%) and validation (5\%) sets.

\paragraph{Lung Cancer Prediction Problem}
\label{paragraph:lung_cancer_biopsy_outcome_prediction}

The final evaluation problem was to predict whether a patient undergoing a lung
biopsy would be diagnosed with lung cancer within the prediction horizon. This
prediction covers the case where the biopsy was immediately positive and
treatment begun, as well as the case where the biopsy might be immediately
negative, but lung cancer developed at some later point within the prediction horizon.

Records with at least 10 ICD-9 codes of any type and at least 24 months of data
before the cutoff date were considered for the dataset. They all had either an
ICD-9 code from the 33.2 group (\emph{Diagnostic Procedures On Lung And
  Bronchus}) or a procedure code for a lung biopsy. All instances used
the code for the first lung biopsy as the cutoff date.  Instances with at least
two downstream codes with an ICD-9 162 parent (\emph{Malignant neoplasm of
  trachea bronchus and lung}) were labeled as positive, and instances with no
codes from that parent were labeled negative.

Because both positive and negative instances were selected by the presence of a
lung biopsy, matching was not needed to reduce information leaking from the
selection criteria.  The positive cohort contained of 1104 instances, while the
negative cohort contained 5631 instances. A stratified split was performed to
divide the dataset into training (75\%), test (20\%) and validation (5\%) sets.

\subsection{Subjective Evaluation}
\label{subsec:subjective_evaluation}

We subjectively explored the properties of the embedding by visualizing the
embedded patient space in two dimensions. To understand the degree to which
similar patients were placed nearby in the space, we projected the positive
instances, negative instances, and some randomly sampled additional records
into the first two principal components of the embedding space and plotted this
for each of the prediction problems. The principal components were computed
separately for the input dataset of each problem.

And finally, to understand the degree to which the embedding captures the
notion of a disease trajectory in this space, we looked at several longitudinal
trajectories of individual records, overlaying them onto the 2-dimensional
projection.

\section{Results and Discussion}
\label{sec:results_and_discussion}

Objectively, the embedded models were robust to most architecture choices and
parameter settings, and they captured a large fraction of the complex
dependencies in the data, although in some cases the compression did lose
information. Subjectively, the embeddings capture quite well the notions of
patient similarity and disease trajectory.

\subsection{Embedding Model Architecture}
\label{subsec:embedding_models}

For these problems the Distributed Bag of Words Model outperformed the
Distributed Memory Model by a fair amount, but otherwise performance was robust
to changes in embedding model parameters; there was a slight preference for
hierarchical softmax over negative sampling, and no substantial effect of
embedding dimension above about 50 (\cref{models-table}).  For the Distributed
Bag of Words Model, there was no systematic preference for large vs. small
window sizes, although for the Distributed Memory Model smaller windows worked
better.

\subsection{Prediction Tasks}
\label{subsec:prediction_tasks}

The prediction task results demonstrate that the embedding manages to capture
a large fraction of the data dependencies with minimal loss, although the degree of
each of these varied by task (\cref{auc-calibration-figure}).

In the breast cancer prediction, the embedded representation captured as much
complex information as the complex model did, with no apparent information
loss. This is the best we could hope for with the performance of the embedding;
if it were always true, it would mean we could routinely store the complex
dependencies in the data representation, and then use fast, simple linear
models for all of our prediction tasks.

However, in the diabetic treatment prediction, the embedded representation
provided no improvement over the simple representation, indicating its failure
to extract additional information from the data. On the other hand, the small
improvements under the complex model suggest that there was little increase to
be had. In this case, there was only a small amount of additional structure
available in the data to improve the prediction (which was already very
accurate without using any complex interaction information), and this small
additional structure was picked up only by the complex model. And in fact, some
of that small additional dependence information was lost by the embedding.

Finally, in the lung cancer prediction, the embedded representation captured
much additional structure, more than half of what was captured by the complex
model, although there was also some information lost.

\begin{figure}[t]
  \includegraphics[width=3.33in]{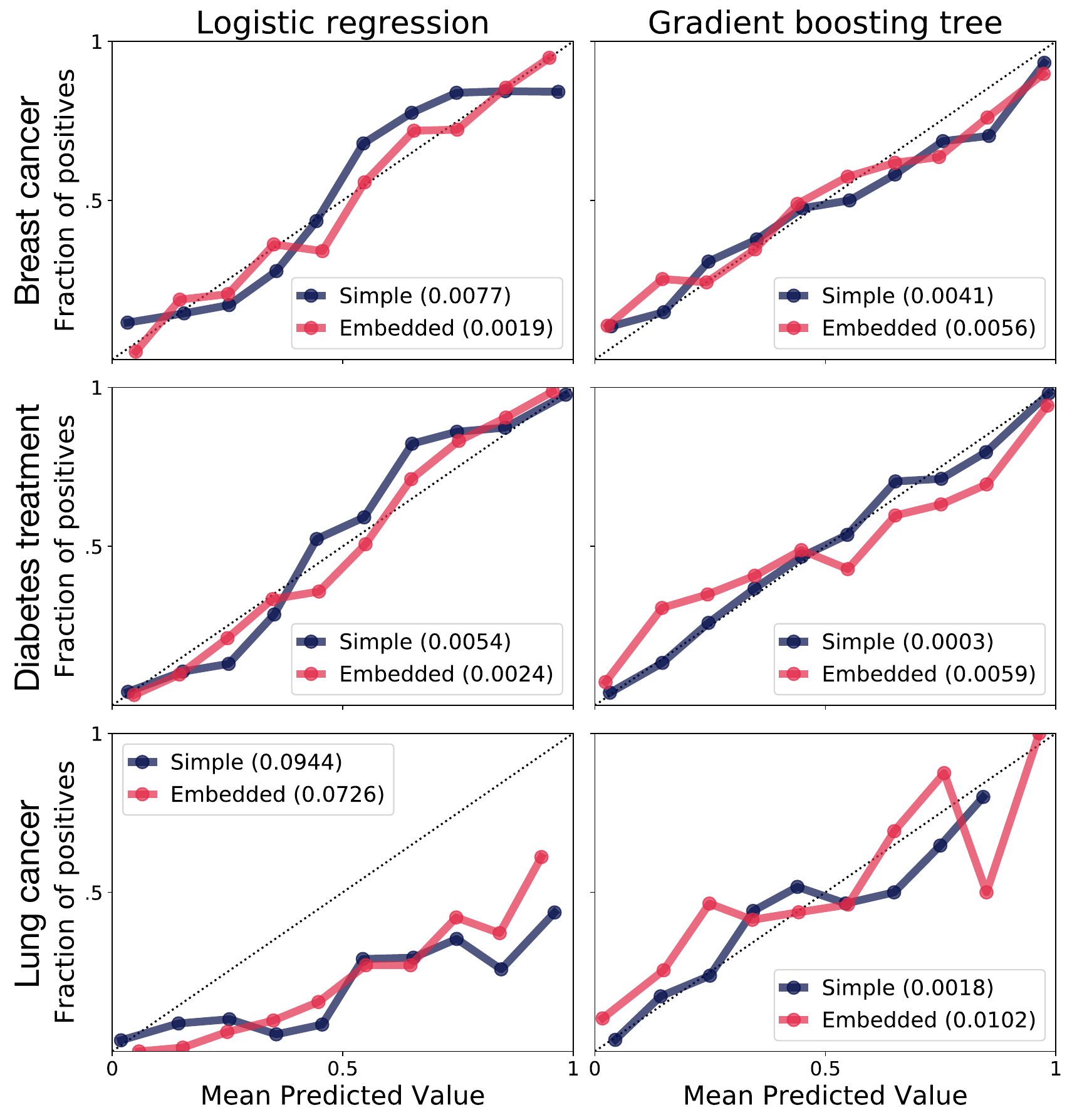}
  \caption{Calibration was largely unaffected by the choice of representation for
    any problem. Dotted line is the expected value for a perfect
    calibration. Value in parentheses is the MSE over all bins of observed
    vs. expected. Data shown is for a 1 day prediction horizon, and is typical
    of other cases.}
  \label{calibration-figure}
\end{figure}

The calibration of predictions was not changed much by the choice of
representation (\cref{calibration-figure}). Of the small differences,
calibration of the linear model was always slightly better for the embedded
representation, and for the complex model it was slightly better for the simple
representation. The calibration for both representations was poor for the
linear model of lung cancer.

\begin{figure*}
  \includegraphics[width=7in]{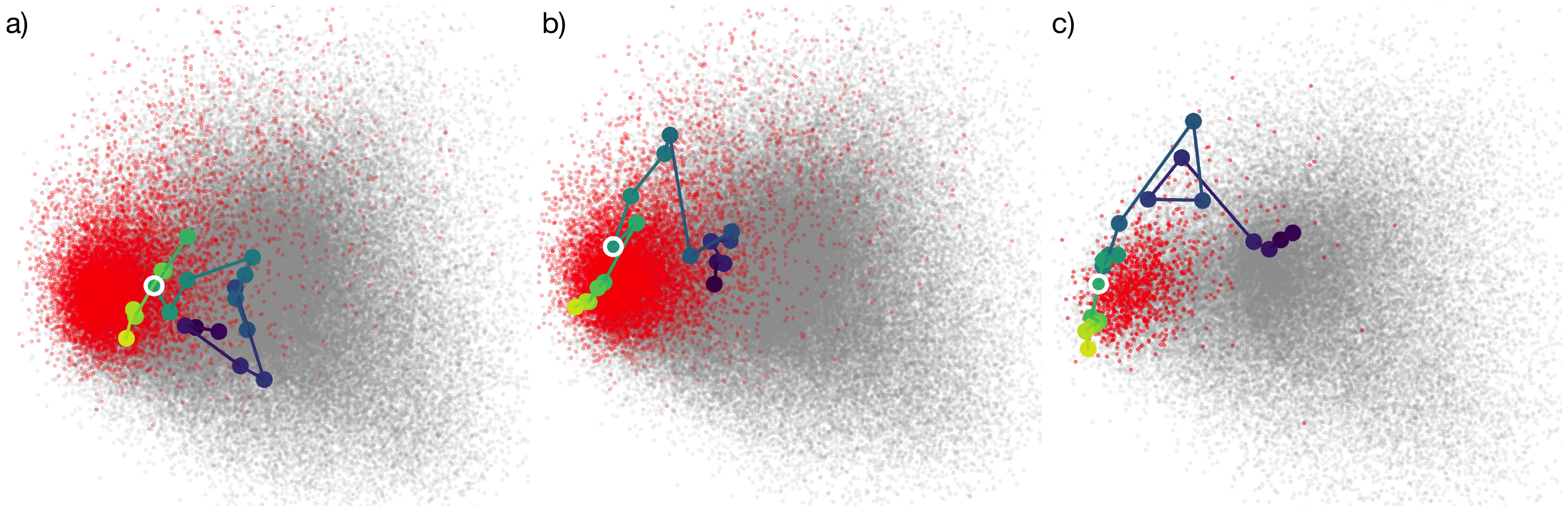}
  \caption{The embedded representation preserves notions of patient similarity
    and disease trajectory for all three problems: a) Breast cancer prediction,
    b) Diabetes treatment prediction, c) Lung cancer prediction. Each small dot
    represents a patient record projected onto the first 2 principal components
    for each dataset. Red dots are positive labeled records, which cluster
    tightly in each case, indicating the preservation of patient
    similarity. Grey dots are negative records plus a large sample of random
    records from the embedding dataset. A single patient trajectory is overlaid
    in each figure, where the passage of time is indicated by a progression of
    dark purple to yellow color. A white circle indicates the time when the
    diagnosis was made or the treatment started. In all cases the patient
    starts out negative and outside the cluster, and progresses to positive
    inside the cluster.}
  \label{trajectory-figure}
\end{figure*}

\subsection{Subjective Analysis}
\label{subsec:subjective_analysis}

The embedded vectors subjectively do a great job of encoding patient state in a
way that enables simple similarity metrics and disease trajectory visualizations
(\cref{trajectory-figure}).  We evaluated a couple dozen of these
visualizations, and we show here a representative case from each problem. Each
panel shows the same embedding space, but projected onto the first two principal
components for the corresponding problem. In each panel the positive instances
are clustered tightly in one area of the projection, although there is some fade
between the positive and negative instances. The fade may be due to undiagnosed
illness, label noise, limitations in the projection, or limitations in the
embedding. The patient trajectories each illustrate the disease progression of a
single patient from having no disease and located among the negative instances,
to having the disease and located among the positive instances. The diagnosis of
the disease or the start of treatment happens as they cross the boundary to the
positive cases. Static and animated GIF images of trajectories for more patients
are available in our code
repository\footnote{https://github.com/ComputationalMedicineLab/patient2vec}.

\subsection{Conclusions}
\label{subsec:conclusions}

This work demonstrates merit in the idea of capturing complex data dependencies
and storing them in the data representation instead of in the prediction model,
especially if the original data is as messy as clinical data. The advantages of
using a record-level semantic embedding for such a representation are that the
complexity can be captured independent of any particular learning problem,
stored much more compactly than the original data, and then used with simple,
fast linear models for prediction.

This type of design may work especially well for initial, proof-of-concept
experiments, such as rough-cut cohort definitions where the trade-off between
accuracy and time-to-result falls preferentially toward getting faster answers.
The generic nature of the representation and the fact that it is computed in an
unsupervised way also lends itself to situations where hundreds or thousands of
approximate models need to be built, such as to predict whether the patient will
develop any of the 18,000 conditions described by ICD-9 codes. There may be more
powerful and accurate ways to train models to do that, but the cost in computing
time and software engineering effort may not be worth it for some use cases.

One way to improve our results may be to add more data types that are found in
a typical EHR. Clinical notes could be added in the usual way for training
semantic embeddings, and other structured data such as vital signs measurements
could also be added. This is the focus of future work.

The method presented here provides a promising rapid research approach for
researchers working with EHR data, as well as knowledge discovery and
exploration of complicated, heterogeneous datasets more broadly.
\section*{Acknowledgments}

This work was funded by grant R01EB020666 from the National Institute of
Biomedical Imaging and Bioengineering. Clinical data was provided by the
Vanderbilt Synthetic Derivative, which is supported by institutional funding
and by the Vanderbilt CTSA grant ULTR000445.

\bibliographystyle{ACM-Reference-Format}
\bibliography{embed-comp}

\end{document}